\documentclass[12pt]{article}
\usepackage{graphics}
\usepackage{epsfig}
\usepackage{dcolumn}% Align table columns on decimal point
\usepackage{bm}% bold math

\newcommand{\bq}{\begin{eqnarray} }
\newcommand{\eq}{\end{eqnarray} }

\begin{document}\begin{titlepage}

\begin{flushright}
%\preprint{
OSU-HEP-05-10\\
July 2005\\
%}
\end{flushright}

\vspace*{1.0cm}

\begin{center}
{\Large {\bf Model of Geometric Neutrino Mixing\\[-0.05in]
}}
\end{center}
\vspace*{0.3cm}
\begin{center}
 { {\bf K.S. Babu$^1$}} and { {\bf Xiao-Gang He$^{2,3}$}}\\
\vspace*{0.3cm}
$^1$Oklahoma Center for High Energy Physics\\
Department of Physics, Oklahoma State University\\
Stillwater, OK~74078, USA\\

\vspace*{0.1in}
 $^2$Department of Physics,
 Nankai University,
  Tianjin, China\\

 \vspace{0.1in}
 $^3$NCTS/PTE, Department of Physics,
  National Taiwan University, Taipei
\end{center}

 \vspace*{0.5cm}

%%%%%%%%%%%%%%%%%%%%%%%%%%%%%%%%%%%%%%%%%%%%%%%%%%%%%%%%%%%%%%%%%%%%%%%%%%%%%%%%%%%%%%%%%%%%%%%%%%%%%%%%%%%%%%%%%%%%%
\begin{abstract}

Current neutrino oscillation data from solar, atmospheric, and
reactor experiments are consistent with the neutrino mixing matrix
elements taking values $\sin^2\theta_{12} = 1/3,~\sin^2\theta_{23}
= 1/2$, and $\sin^2\theta_{13}=0$.  We present a class of
renormalizable  gauge models which realize such a geometric mixing
pattern naturally. These models, which are based on the
non--Abelian discrete symmetry $A_4$, place significant
restrictions on the neutrino mass spectrum, which we analyze.  It
is shown that baryogenesis via leptogenesis occurs quite
naturally, with a single phase (determined from neutrino
oscillation data) appearing in leptonic asymmetry and in
neutrinoless double beta decay. Such predicted correlations would
provide further tests of this class of models.

\end{abstract}

\end{titlepage}
\newpage

\section*{Introduction}

Our understanding of the fundamental properties of neutrinos has
improved dramatically over the last few years. Atmospheric and
solar neutrino experiments have by now firmly established
occurrences of  neutrino flavor oscillations \cite{data}.  In
order for neutrinos to oscillate, they must have non--degenerate
masses. In addition, different neutrino flavor states must mix
with one another. When positive evidence for oscillations from
solar and atmospheric neutrinos are combined with results from
reactor data \cite{analysis,lisi}, one obtains the following
neutrino mass and mixing pattern (with 2$\sigma$ error bars)
\cite{lisi}:
\begin{eqnarray}
&&\Delta m^2_{\odot} = m_2^2-m_1^2 = 7.92\times 10^{-5} (1\pm
0.09)\;\mbox{eV}^2,\\
&&\Delta m^2_{\rm atm} = m_3^2-m_2^2 =
\pm 2.4\times 10^{-3} (1^{+0.21}_{-0.61})\;\mbox{eV}^2,\\
&&\sin^2\theta_{12} = 0.314
(1^{+0.18}_{-0.15})\;,\;\;\sin^2\theta_{23} = 0.44
(1^{+0.41}_{-0.22})\;,\;\;\sin^2\theta_{13} =
0.9^{+0.23}_{-0.9}\times 10^{-2}~.
\end{eqnarray}
Here $m_i$ are the (positive) neutrino mass eigenvalues, and $\theta_{ij}$ are the neutrino mixing angles.
$m_2^2 - m_1^2 > 0$ in Eq. (1) is necessary for MSW resonance to occur inside the Sun.  The sign of
$\Delta m^2_{\rm atm}$, which is physical,  is currently unknown.

A remarkable feature of the oscillation data is that they are all
consistent with a ``geometric" neutrino mixing pattern defined by
the parameters $\sin^2\theta_{12} = 1/3$, $\sin^2\theta_{23} =
1/2$, and $\sin^2\theta_{13} = 0$.  In fact, these geometric
mixing angles  are very close to the central values of Eq. (3). We
observe that unlike the quark mixing angles, which are related to
the quark mass ratios in many models (eg: $\theta_C \simeq
\sqrt{m_d/m_s}$), the neutrino mixing angles seem to be unrelated
to the neutrino mass ratios.

The purpose of this Letter is to provide a derivation of such a
geometric neutrino mixing based on renormalizable gauge theories.
The neutrino mixing matrix (the MNS matrix) that we will derive
has the form \cite{hps,xing,he,wolf}
\begin{eqnarray}
U_{MNS} = \left( \matrix{\sqrt{{2 \over 3}} & {1 \over \sqrt{3}} &
0 \cr -{1 \over \sqrt{6}} & {1 \over \sqrt{3}} & -{1 \over
\sqrt{2}} \cr -{1 \over \sqrt{6}} & {1 \over \sqrt{3}} & {1 \over
\sqrt{2}} } \right)P~. \label{geometric}
\end{eqnarray}
Here $P$ is a diagonal phase matrix which is irrelevant for
neutrino oscillations, but relevant for neutrinoless double beta
decay. Eq. (\ref{geometric}) yields the desired values,
$|U_{e2}|^2 = 1/3,~|U_{\mu 3}|^2 = 1/2,~ |U_{e3}|^2 = 0$.

Our derivation of Eq. (\ref{geometric}) will be based on the
non--Abelian discrete symmetry $A_4$, the symmetry group of a
regular tetrahedron.  This symmetry group has found application in
obtaining maximal atmospheric neutrino mixing \cite{mar} and in
realizing quasi--degenerate neutrino mass spectrum \cite{bmv}. No
successful derivation of Eq. (\ref{geometric}) has been achieved
to our knowledge (based on $A_4$ or other symmetries) in a
renormalizable gauge theory context. For attempts along this line
see Ref. \cite{hps,af,man}.  In Refs. \cite{hps,he}, Eq.
(\ref{geometric}) was suggested as a phenomenological  ansatz. In
Ref. \cite{af}, a higher dimensional set up is used to motivate
Eq. (\ref{geometric}).  Ref. \cite{man} analyzes special cases of
an $A_4$ derived neutrino mass matrix towards obtaining the
structure of Eq. (\ref{geometric}). A large number of models in
the literature have derived maximal atmospheric mixing based on
non--Abelian symmetries \cite{discrete}, but in most models the
solar mixing angle is either maximal (now excluded by data) or is
a free parameter.

We will see that the derivation of Eq. (\ref{geometric}) places
strong restrictions on the neutrino mass pattern. We find that the
out of equilibrium decay of the lightest right--handed neutrino
generates lepton asymmetry at the right level to explain the
observed baryon asymmetry of the universe.  A single phase appears
in leptonic asymmetry as well as in neutrinoless double beta
decay, thus providing some hope for testing high scale phenomena
via low energy experiments.

\section*{The Model}

We work in the context of low energy supersymmmetry, which is
motivated by a solution to the gauge hierarchy problem as well as
by the observed unification of gauge couplings. The gauge group of
our model is that of the Standard Model, $SU(3)_C \times
SU(2)_L\times U(1)_Y$. We augment this symmetry with a
non--Abelian discrete symmetry $A_4$. This order 12 group is  the
symmetry group of a regular tetrahedron.  $A_4$ has a unique
feature in describing the lepton sector: It has one triplet and
three inequivalent singlet representations, thus allowing for
assigning the left--handed lepton fields to the triplet and the
right--handed charged lepton fields to the three inequivalent
singlets.

Denoting the three singlets of $A_4$ as $({\bf 1,~1',~1''})$, with
the ${\bf 1}$ being the identity representation, we have ${\bf 1'}
\times {\bf 1''} = {\bf 1}$, ${\bf 1'} \times {\bf 1'} = {\bf
1''}$, and ${\bf 1''} \times {\bf 1''} = {\bf 1'}$. Furthermore,
${\bf 3} \times {\bf 3} = {\bf 1} + {\bf 1'} + {\bf 1''} + {\bf
3}_s + {\bf 3}_a$. Specifically, for the product of two triplets
we have $(a_1, a_2, a_3) \times (b_1, b_2,b_3) = (a_1 b_1 + a_2
b_2 + a_3 b_3)~ ({\bf 1})$;  $(a_1 b_1 + \omega^2 a_2 b_2 + \omega
a_3 b_3) ~({\bf 1'})$; $(a_1 b_1 + \omega a_2 b_2 + \omega^2 a_3
b_3) ~({\bf 1''})$; $(a_2 b_3 + a_3 b_2, a_3 b_1+ a_1 b_3, a_1 b_2
+ a_2 b_1)~({\bf 3}_s$).  Here $\omega = e^{2 i \pi/3}$.

In addition to the $A_4$ symmetry, we assume a $Z_4 \times Z_3$
discrete symmetry.  The $Z_4$ is an $R$--symmetry  under which the
superpotential carries  2 units of charge.  The $Z_4$ and $Z_3$
symmetries are broken softly in the superpotential via the lowest
dimensional operators.

The lepton and Higgs fields transform under $A_4\times Z_4\times
Z_3$ as follows.
\begin{eqnarray}
&&L: (3,1,0),\;\;e^c: (1+1'+1'',3,0),\;\;\nu^c: (3,0,1),\;\;E: (3,1,0),\;\;E^c: (3,1,0),\nonumber\\
&&H_u: (1,1,2),\;\;H_d:(1,0,0),\;\;\chi: (3,2,0),\;\;\chi':
(3,2,1),\;\;S_{1,2}: (1,2,1).
\end{eqnarray}
Here in the fermion sector we have introduced new vector--like
iso--singlet fields $E$ and $E^c$ transforming under the SM gauge
group as (1,1,-1) and (1,1,1), respectively, which will acquire
large masses and decouple. $H_u$ and $H_d$ are the usual Higgs
fields of MSSM, while $\chi, \chi', S_{1,2}$ are all SM singlet
fields needed for achieving symmetry breaking. The quark fields
$(Q, u^c, d^c)$ are all singlets of $A_4$ with $Z_4 \times Z_3$
charges of $Q(1,1); u^c(0,0)$ and $d^c(1,2)$, so that the usual
quark Yukawa couplings $Q d^c H_d + Q u^c H_u$ are allowed in the
superpotential.\footnote{A bare mass term $\mu H_u H_d$ is not
allowed in the superpotential by the symmetries, but the Kahler
potential, which is assumed to not respect these symmetries,
allows a Planck mass suppressed term, ${\cal L} \supset \int H_u
H_d Z^* d^4 \theta/M_{\rm Pl}$, generating the required $\mu$
term.}

The superpotential terms relevant for lepton masses consistent
with the symmetries is
\begin{eqnarray}
W_{\rm Yuk} &=& M_E E_i E^c_i + f_e L_i E^c_i H_d + h^e_{ijk}E_i
e^c_j \chi_k+ {1\over 2} f_{S_l} \nu^c_i \nu^c_i S_l + {1\over
2}f_{ijk} \nu^c_i\nu^c_j \chi'_k +  f_\nu L_i \nu^c_i
H_u.\label{yukawa}
\end{eqnarray}
Here the flavor structure of the three independent $h^e_{ijk}$
couplings and the one independent $f_{ijk}$ coupling can be easily
obtained from the $A_4$ multiplication rules given earlier.

The Higgs superpotential of the model is given by
\begin{eqnarray}
W_{\rm Higs} &=& \lambda_\chi \chi_1\chi_2\chi_3 + \lambda_{\chi'
s}(\chi_1'^{2}+\chi'^2_2+\chi'^2_3) S_1 +\lambda'_{\chi'}
\chi'_1\chi'_2\chi'_3\nonumber\\
& +& \lambda_{s_{11}} S^3_1 +\lambda_{s_{12}}S^2_1 S_2 +
\lambda_{s_{21}}S_1S^2_2 + \lambda_{s_{22}}S^3_2 \nonumber
\\
&+&\mu^2_1 S_1 + \mu^2_2 S_2 +\mu_\chi
(\chi_1^2+\chi^2_2+\chi^2_3).\label{potential}
\end{eqnarray}
Here $\chi = (\chi_1, \chi_2, \chi_3)$, and $\chi' = (\chi_1',
\chi_2', \chi_3')$.  The last three terms in the last line of Eq.
(\ref{potential}) break the $Z_4$ and $Z_3$ symmetries softly. The
$\mu_{1,2}^2$ terms are the lowest dimensional terms that break
the $Z_3$ symmetry softly, while leaving $Z_4$ unbroken. The
$\mu_\chi$ term is the lowest dimensional term that breaks the
$Z_4$ symmetry softly. Such soft breaking can be understood as
spontaneous breaking occurring at a higher scale. We have chosen
without loss of generality the combination of $S_1$ and $S_2$ that
couples to $\chi'$ as simply $S_1$ in Eq. (\ref{potential}).

Minimizing the potential derived from Eq. (\ref{potential}) in the
supersymmetric limit, we obtain the following vacuum structure:
\begin{eqnarray}
\left\langle S_2 \right\rangle = v_s,\;\; \left\langle S_1
\right\rangle = 0;\;\; \left\langle \chi_1 \right\rangle =
\left\langle \chi_2\right \rangle = \left\langle
\chi_3\right\rangle  = v_\chi;\;\;\left\langle
\chi'_2\right\rangle = v_{\chi'},\;\;\left\langle
\chi_1'\right\rangle = 0;\;\; \left\langle \chi_3' \right\rangle =
0.\label{vev}
\end{eqnarray}
with $v_\chi = - 2 \mu_\chi/\lambda_\chi$, $v_s^2 = - \mu_2^2/(3
\lambda_{s_{22}})$, and $v_{\chi'}^2 = (\lambda_{s_{21}}\mu_2^2 -
3 \lambda_{s_{22}}\mu_1^2)/(3 \lambda_{s_{22}}\lambda_{\chi's})$.
Electroweak symmetry breaking is achieved in the usual way by
$\left\langle H_u \right\rangle = v_u$, $\left\langle H_d
\right\rangle =v_d$. We emphasize that vanishing of certain VEVs
is a stable result, owing to the discrete symmetries present in
the model. This is important for deriving the MNS matrix of Eq.
(\ref{geometric}). We observe that there are no pseudo-Goldstone
modes, as can be seen by directly computing the masses of the
Higsinos from Eq. (\ref{potential}).

The mass matrices $M_{eE}$ for the charged leptons and $M_{\nu
\nu^c}$ for the neutral leptons resulting from Eqs. (\ref{yukawa})
and (\ref{vev}) are given by (in the notation ${\cal L} = (e, E)~
M_{eE}~(e^c, E^c)^T$)
\begin{eqnarray}
&&M_{eE} = \left ( \begin{array}{llllll}
0&0&0&f_e v_d&0&0\\
0&0&0&0&f_e v_d&0\\
0&0&0&0&0&f_e v_d\\
h^e_1 v_\chi &h^e_2v_\chi& h^e_3 v_\chi&M_E&0&0\\
h^e_1 v_\chi &h^e_2\omega v_\chi& h^e_3 \omega^2 v_\chi&0&M_E&0\\
 h^e_1 v_\chi
&h^e_2\omega^2v_\chi& h^e_3 \omega v_\chi&0&0&M_E
\end{array}
\right )\;,\nonumber\\
\nonumber\\
 &&M_{\nu \nu^c} = \left ( \begin{array}{llllll}
0&0&0&f_\nu v_u&0&0\\
0&0&0&0&f_\nu v_u&0\\
0&0&0&0&0&f_\nu v_u\\
f_\nu v_u&0&0&f_{s_2}v_s&0&f_{\chi'}v_{\chi'}\\
0&f_\nu v_u&0&0&f_{s_2} v_s&0\\
0&0&f_\nu v_u&f_{\chi'}v_{\chi'}&0&f_{s_2}v_s
\end{array}
\right ).\label{mr}
\end{eqnarray}

Since the $E$ and the $E^c$ fields acquire large masses, of order
the GUT scale, they can be readily integrated out. The reduced
$3\times 3$ mass matrices for the light charged leptons is given
by
\begin{eqnarray}
&&M_{e} = U_L \left ( \begin{array}{lll}
m_e &0&0\\
0&m_\mu&0\\
0&0&m_\tau
\end{array}
\right ),\;\; U_L = {1\over
\sqrt{3}}\left ( \begin{array}{lll} 1&1&1\\
1&\omega&\omega^2\\
1&\omega^2&\omega
\end{array}
\right ),
\end{eqnarray}
where $m_i = \sqrt{3}(f_e v_d v_\chi/M_E) h^e_i(1+(h_i
v_\chi)^2)/M_E^2)^{-1/2}$.  The light neutrino mass matrix is
found to be
\begin{eqnarray}
M^{\rm light}_\nu = m_0 \left (\begin{array}{ccc}
1&0&x\\
0&1-x^2& 0\\
x&0&1
\end{array}
\right ),
\end{eqnarray}
where $m_0 = f^2_\nu v_u^2 f_{s_2}v_s/(f^2_{s_2}v_s^2 -
f^2_{\chi'}v^2_{\chi'})$, and $x=- f_{\chi'}v_{\chi'}/(f_{s_2}
v_s)$.   We define $x = |x| e^{i\psi}$.

$M_\nu^{\rm light}$ can be diagonalized by the transformation
$M_\nu^{\rm light} = U_\nu^* D_\nu U_\nu^\dagger$ with
\begin{eqnarray}
U_\nu = { 1 \over \sqrt{2}} \left(\matrix{1 & 0 & -1 \cr 0 &
\sqrt{2} & 0 \cr 1 & 0 & 1 }\right)P^*; ~~~D_\nu = m_0
\left(\matrix{ |1+x| & ~ & ~ \cr ~ & |1-x^2| & ~ \cr ~ & ~ &
|1-x|}\right). \label{mass}
\end{eqnarray}
$P^*$ is a diagonal phase matrix given by
\begin{eqnarray}
P^*=diag\{e^{-i\phi_1/2},~e^{-i(\phi_1+\phi_2)/2},~e^{-i(\phi_2+\pi)/2}\}
,\;\; \phi_1 = arg(1+x),~\phi_2=arg(1-x)~.
\end{eqnarray}
These Majorana phases will not be relevant for neutrino
oscillations, but they will appear in neutrinoless double beta
deacy and in leptogenesis. The MNS matricx is given by $U_{MNS} =
U_L^TU_\nu^*$ which has the form given in Eq. (\ref{geometric}).
(In making this identification, we make a field redefinition of
the left--handed charged lepton fields, $e_L = diag.\{1, \omega,
\omega^2\}e_L'$.)  This is the main result of this paper.

\section*{Constraints on neutrino masses}

From Eq. (\ref{mass}), the expressions for the mass eigenvalues
can be inverted to obtain the following relations for the
parameters $|m_0|,~|x|$ and $\psi = arg(x)$:
\begin{eqnarray}
|m_0| &=& {m_1 m_3 \over m_2} \nonumber \\
|x| &=& { 1 \over {\sqrt{2} m_1 m_3}} \left[m_1^2 m_2^2 + m_2^2
m_3^2 - 2 m_1^2 m_3^2\right]^{1/2},\nonumber\\
 \cos\psi &=&
{-(m_3^2 - m_1^2) m_2^2 \over  {2 \sqrt{2} m_1 m_3 \left[m_1^2
m_2^2 + m_2^2 m_3^2 - 2 m_1^2 m_3^2\right]^{1/2}}}. \label{masses}
\end{eqnarray}
Here $m_1 = |m_0(1+x)|, ~m_2= |m_0(1-x^2)|,~m_3=|m_0(1-x)|$. There
are restrictions arising from the conditions that $|x|$ be real
and $|\cos\psi| \leq 1$, which we analyze now.

Because of the observed hierarchy $\Delta m^2_{\rm atm} \gg \Delta
m^2_{\odot}$,  and the requirement of MSW resonance for solar
neutrinos, two possible neutrino mass ordering are allowed. (i)
$m_1 < m_2 < m_3$ (normal mass ordering) and (ii) $m_3 < m_1 <
m_2$ (inverted mass ordering).

If the neutrino masses are strongly hierarchical, $m_1 \ll m_2 \ll
m_3$, then from Eq. (\ref{masses}) one sees that $|\cos\psi| \leq
1$ cannot be satisfied, since $|\cos\psi| \simeq |m_2/(2 \sqrt{2}
m_1)| \gg 1$ in this case.  Similarly $m_3 \ll m_1 \ll m_2$ is
also not allowed. We find that only two possibilities can arise,
depending on (a) $m_2^2-m_1^2 \sim m_2^2+ m_1^2$ (normal
ordering), and (b) $m_2^2 -m_1^2 \ll m_2^2 + m_1^2$ (inverted
ordering). We consider these cases in turn.

These conclusions can also be arrived at by analyzing the neutrino
mass matrix in the flavor basis, i.e., in a basis where the
charged lepton mass matrix is diagonal:
\begin{eqnarray}
M_\nu^{\rm flavor} = {m_0\over 3} \left (
\begin{array}{ccc}
3+2 x- x^2&- x- x^2&- x- x^2\\
- x- x^2&2 x- x^2&3- x- x^2\\
- x- x^2&3- x- x^2&2 x- x^2
\end{array}
\right ).\label{charge}
\end{eqnarray}
When $x = -1 + q$ with $|q| \ll 1$, we see that entries in the
first row and column of Eq. (\ref{charge}) become small.  This
will be the case of normal ordering of masses.

\subsection*{(a) Normal ordering}

This case is realized when $m_2^2 - m_1^2 \sim m_2^2 + m_1^2$. The
condition $|\cos\psi|\leq 1$ can be satisfied only if $m_1/m_2
\simeq 1/2$. Making expansions in small $m_1/m_3$ and small $(2
m_1/m_2-1)$, we find
\begin{eqnarray}
|x| &\simeq& 1 + {m_1^2 \over m_3^2} -2\left({2 m_1 \over m_2}-1\right), \nonumber \\
\cos\psi &\simeq& -\left[1-2\left\{{m_1^2 \over m_3^2} - \left({2 m_1 \over m_2} - 1\right)^2\right\}\right]~.
\end{eqnarray}
We see a further restriction that $m_1/m_3 \geq |2m_1/m_2-1|$.

Neutrinoless double beta decay is sensitive to the effective mass
\begin{equation}
m_{\beta \beta} = \left|\sum_iU_{ei}^2 m_i\right| = {|m_0| \over
3} |3-x||1+x|.
\end{equation}
In the case under study this takes the value
\begin{equation}
m_{\beta \beta} \simeq { 4 \over 3}m_1 \simeq {4 \over 3} \left
({\Delta m^2_{\odot}\over 3}\right)^{1/2} \simeq 0.0068 ~{\rm
eV}~.
\end{equation}
The effective neutrino mass measurable in tritium beta decay $m_{\nu_e}$ is given by
\begin{equation}
m_{\nu_e} = \left[\sum_i|U_{ei}|^2 m_i^2\right]^{1/2}
\end{equation}
which in this case takes the value
\begin{equation}
m_{\nu_e} \simeq \left[{2 \over 3} \Delta m^2_{\odot}\right]^{1/2}
\simeq 0.0073 ~{\rm eV}.
\end{equation}
Here we made use of the fact that $3m_1^2 \simeq \Delta
m^2_{\odot}$. Finally, the sum of neutrino masses, sensitive to
cosmological measurements, is given by
\begin{equation}
\sum_i m_i \simeq 3 m_1+ m_3 \simeq  3 \left ( {\Delta
m^2_{\odot}\over 3} \right )^{1/2}+ \left|\Delta m^2_{\rm
atm}\right|^{1/2} \simeq 0.064 ~{\rm eV}~.
\end{equation}

\begin{figure}[htb]
\begin{center}
\begin{tabular}{c}
\hspace*{-1.5cm}
\includegraphics[width=16cm]{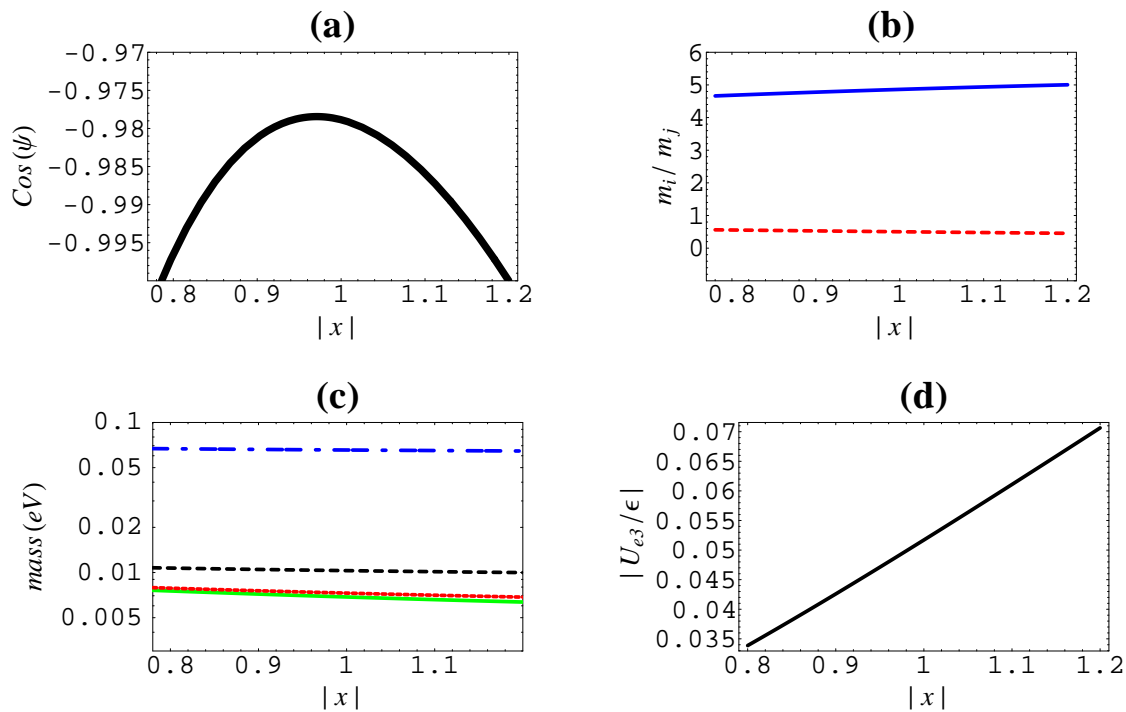}
\end{tabular}
\end{center}
\caption{Various quantities as functions of $|x|$ for normal mass
hierarchy case. (a) $\cos\psi$ vs. $|x|$; (b) $m_1/m_2$ (dashed)
and $m_3/m_2$ (solid) vs. $|x|$; (c) $m_{ee}$ (solid (green)), $
m_{\nu_e}$ (dotted (red)), $m_2$ (dashed (black)) and $\sum m_i$
(dot-dashed (blue)) (in eV unit) vs. $|x|$; (d) RG running
correction to $|U_{e3}/\epsilon|$ vs. $|x|$.}\label{Mi}
\label{obsevables}
\end{figure}

We have plotted in Fig. 1a - 1c various masses and mass ratios as
functions of $|x|$ using the exact expressions with central values
for $\Delta m^2_\odot$ and $\Delta m^2_{atm}$. These results
confirm our analytical solutions.

\subsection*{(b) Inverted mass ordering}

In this case, $m_2^2 - m_1^2 \ll m_2^2 + m_1^2$.  The small
parameter expansion is different from (a), since the denominator
of Eq. (16) in $\cos\psi$ becomes small.  Here $m_1$ and $m_2$ are
nearly equal, and $m_3$  will turn out to be much smaller than
$m_1$. In order to satisfy $\Delta m^2_{\odot} \ll \Delta m^2_{\rm
atm}$ it is necessary that $\cos\psi \simeq |x|/2$, which then
requires $|x| \leq 2$.  Writing
\begin{equation}
\cos\psi = {|x| \over 2}(1+q),~q \ll 1,
\end{equation}
we find
\begin{eqnarray}
\Delta m^2_{\odot} &=& m_2^2 - m_1^2 \simeq -|m_0|^2 |x|^2(1+2|x|^2) q \nonumber \\
\Delta m^2_{\rm atm} &=& m_3^2 - m_2^2 \simeq -2|m_0|^2 |x|^2~.
\end{eqnarray}
It becomes clear that $m_3^2 < m_2^2$ (and thus $m_3^2 < m_1^2$),
that is, this case corresponds to an inverted hierarchy.  With
negative $q \ll 1$, the hierarchy in the two oscillation
parameters can be accommodated.

In this case the effective mass for double beta decay is given by
\begin{equation}
m_{\beta \beta } \simeq {m_3 \over 3} \left[9 + { \Delta m^2_{\rm atm} \over m_3^2}\right]^{1/2}
\left[1- {  \Delta m^2_{\rm atm} \over m_3^2}\right]^{1/2}
\end{equation}
with $\Delta m^2_{\rm atm}$ being negative. Here we used the fact
that $|m_0| \simeq m_3$.  The value of $m_3$ is not determined by
oscillation data.  If $m_3^2 \gg |\Delta m^2_{\rm atm}|$, we have
three--fold degeneracy of masses and $m_{\beta \beta}\simeq m_3$.
As $m_3^2 \rightarrow (1/9) |\Delta m^2_{\rm atm}|$, the double
beta decay amplitude vanishes.  Furthermore we have for this case
\begin{equation}
m_{\nu_e} \simeq m_1 \simeq \sqrt{|\Delta m^2_{\rm atm}|
+m_3^2},~~~\sum_i m_i \simeq 2 m_1 + m_3 \simeq 2\sqrt{|\Delta
m^2_{\rm atm}|+m_3^2} + m_3.
\end{equation}

The three--fold degenerate case is obtained from this case by
setting $m_3^2$ much larger than $|\Delta m^2_{\rm atm}|$ (or
equivalently, $|x| \ll 1$).  This case also coincides with the
leading results of Ref. \cite{bmv}.

\begin{figure}[htb]
\begin{center}
\begin{tabular}{c}
\hspace*{-1.5cm}
\includegraphics[width=16cm]{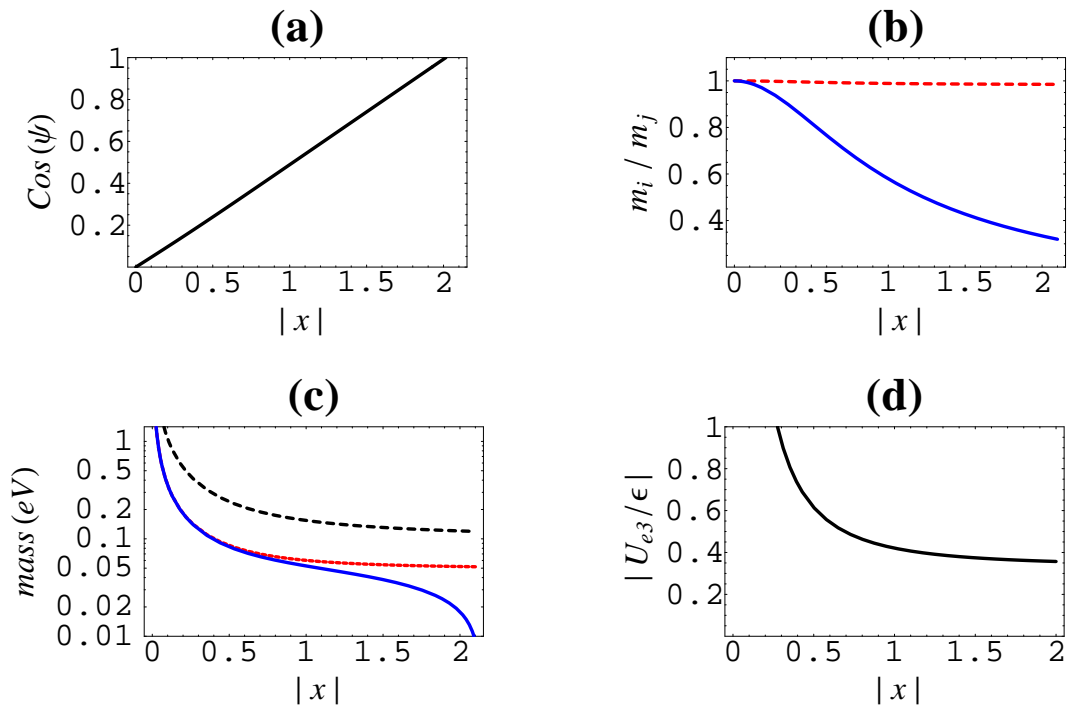}
\end{tabular}
\end{center}
\caption{Various quantities as functions of $|x|$ for inverted
mass hierarchy case. (a) $\cos\psi$ vs. $|x|$; (b) $m_1/m_2$
(dashed) and $m_3/m_2$ (solid) vs. $|x|$; (c) $m_{ee}$ (solid
(blue)), $ m_{\nu_e}\simeq m_2$ (dashed (red)) and $\sum m_i$
(dot-dashed (black)) (in eV unit) vs. $|x|$; (d) RG running
correction to $|U_{e3}/\epsilon|$ vs. $|x|$.}\label{Mi}
\label{obsevables1}
\end{figure}

The exact results for various masses and mass ratios are plotted
in Fig. 2a - 2c as functions of $|x|$.  These results confirm the
analytical approximations presented here.

\section*{Stability of $U_{e3}$}

A distinctive feature of the geometric mixing pattern is that
$U_{e3} =0$ at the scale of $A_4$ symmetry breaking, which we have
taken to be of order the GUT scale.  When running from this high
scale to low energy scale ($M_{EW}$), the mixing matrix may
change, in particular $U_{e3}$ may not be zero any more.  One
should ensure that the pattern of Eq. (\ref{geometric}) is not
destabilized, which can happen if the induced $U_{e3}$ is too
large. We demonstrate this stability now.

The leading flavor-dependent effect of the running from high scale
to low scale is given by the one--loop RGE \cite{babu}
\begin{eqnarray}
{dM^e_\nu\over d\ln t} = {1 \over 32 \pi^2}[M^e_\nu Y^\dagger_e
Y_e + (Y^\dagger_e Y_e)^T M^e_\nu]+...
\end{eqnarray}

This leads to correction, to the leading order, to the entries
$M_{13,23}(1-\epsilon)$ and $M_{33}(1-2\epsilon)$ with $\epsilon
\simeq  Y^2_\tau \ln(M_{\rm GUT}/M_{EW})/32\pi^2$. One obtains to
order $\epsilon$,
\begin{eqnarray}
|U_{e3}| \approx {|\epsilon x| \left||x|+\cos\psi +i
\sin\psi\right| \over 3 \sqrt{2}|\cos\psi(|x|+2\cos\psi)| }.
\end{eqnarray}
One obtains the induced $|U_{e3}|$ for the normal and inverted
hierarchies by inserting the corresponding expressions for
$\cos\psi$ and $|x|$ given earlier.

The results  are shown in Figs 1d (normal mass ordering) and in 2d
(inverted ordering) where we plot $|U_{e3}/\epsilon|$ as a
function of $|x|$. We see that the induced $|U_{e3}|$ is small,
too small to be measured by near future experiments for the normal
mass hierarchy case in the whole allowed $|x|$ range. For the
inverted mass hierarchy case for $|x|$ larger than about 0.2,
$|U_{e3}|$ remains small. For smaller values of $|x|$, with
$\epsilon$ of order one (corresponding to $Y_\tau \approx 1$),
$|U_{e3}|$ can be as large as $0.1$ which may be measured in the
future.  In this case, all three neutrinos are nearly degenerate
and the cosmological mass limit on neutrinos will be nearly
saturated. We conclude that the structure of the mixing matrix
derived is not upset by radiative corrections.

\subsection*{Leptogenesis}

Leptogenesis occurs in a simple way in this model via the decay of
the right-handed neutrinos \cite{fy}. The heavy Majorana mass
matrix of $\nu^c$ is given in the model as (see Eq. (\ref{mr}))
\begin{eqnarray}
M_{\nu^c} = M_R \left[\matrix{1 & 0 &-x \cr 0 & 1 & 0 \cr -x & 0 & 1}\right]~.
\end{eqnarray}
The Dirac neutrino Yukawa coupling matrix is proportional to an
identity matrix at the scale of $A_4$ symmetry breaking which we
take to be near the GUT scale.  The $\nu^c$ fields will remain
light below that scale, down to the scale $M_R$. Renormalization
group effects in the momentum range $M_R < \mu < M_{\rm GUT}$ will
induce non-universal corrections to the Dirac neutrino Yukawa
coupling matrix.  Without such non-universality  no lepton
asymmetry will be induced in the decay of right--handed neutrinos.
The effective theory in this momentum range is the MSSM with the
$\nu^c$ fields.

From the renormalization group equation
\begin{equation}
{dY_\nu \over dt} = { 1 \over 16 \pi^2}Y_\nu(Y_\ell^\dagger
Y_\ell) + ...
\end{equation}
where $W = e Y_\ell L H_d + \nu^c Y_\nu L H_u + ..$, we obtain at
the scale $M_R$, $Y_\nu = Y_\nu^0 \times diag(1, 1, 1-\delta$)
with $\delta \simeq (Y_\tau^2 /16 \pi^2) {\rm ln}(M_{\rm
GUT}/M_R)$. $Y_\nu^0$ is the value of the universal Dirac Yukawa
coupling at the GUT scale.

We diagonalize $M_{\nu^c}$ by the rotation $\nu^c = O Q N$, where
\begin{eqnarray}
O = {1 \over \sqrt{2}} \left(\matrix{1 & 0 & 1 \cr 0 & \sqrt{2} & 0  \cr -1 & 0 & 1}\right),~~
Q = diag \{e^{-i\phi_1/2}, ~1,~e^{-i\phi_2/2}\}
\end{eqnarray}
so that the $N$ fields are the mass eigenstates with  real and
positive mass eigenvalues: $ M_{N} = M_R  \times diag(|1+x|, 1,
|1-x|)$.

 In
the basis where the heavy $\nu^c$ fields have been diagonalized,
the Dirac neutrino Yukawa coupling matrix takes the form
$\hat{Y}_\nu = Q O^T Y_\nu$, so that
\begin{eqnarray}
\hat{Y}_\nu \hat{Y}_\nu^{\dagger} = {|Y_\nu^0|^2 \over 2} \left[\matrix{1+(1-\delta)^2 & 0 & -e^{i(
\phi_2-\phi_1)/2} \{(1-\delta)^2-1\} \cr 0 & 2 & 0 \cr -e^{i(\phi_1-\phi_2)/2} \{(1-\delta)^2-1\}
& 0 & 1+(1-\delta)^2}\right]~.
\end{eqnarray}

The CP asymmetry arising from the decay of the field $N_i$ is given by
\begin{equation}
\epsilon_i = {-1 \over 8 \pi} {1 \over [\hat{Y}_\nu \hat{Y}_\nu^\dagger]_{ii}} \sum_j Im\{
[\hat{Y}_\nu \hat{Y}_\nu^\dagger]_{ij}^2\} f\left({M_j^2 \over M_i^2}\right)
\end{equation}
where
\begin{equation}
f(y) = \sqrt{y}\left({2 \over y-1} + log {1+y \over y}\right)~.
\end{equation}
As $y \gg 1$, $f(y) \rightarrow 3/\sqrt{y}$.

In the normal hierarchy case, $m_1/m_3 = M_1/M_3$, so the lightest $N$ field is $N_1$.
In this case we have
\begin{eqnarray}
\epsilon_1 &=& {-3 |Y_\nu^0|^2 \over 8 \pi} \delta^2 \left({m_1 \over m_3}\right) \sin (\phi_2-\phi_1)\nonumber \\
&\simeq& {\pm 3 |Y_\nu^0|^2\over 8 \pi}  \delta^2 \left({m_1 \over
m_3}\right) \left[1 - {(2m_1/m_2-1)^2 \over (m_1/m_3)^2}
\right]^{1/2}~.
\end{eqnarray}
To see the numerical value of $\epsilon_1$, we note that
$|Y_\nu^0|$ can be of order one, $\delta \simeq (0.1 Y_\tau^2)$,
and $m_1/m_3 \simeq [{\Delta m^2_{\rm solar} / 3 \Delta m^2_{\rm
atm}}]^{1/2} \simeq 0.1$. For very large value of $\tan\beta$,
$Y_\tau \simeq 1$, and we find $\epsilon_1 \simeq 10^{-4}$.  Even
for moderate values of $\tan\beta \sim 20$, we find that
$\epsilon_1 \simeq 10^{-6}$ is possible.  The negative sign will
also ensure the correct sign of baryon asymmetry.  The induced
lepton asymmetry is converted to baryon asymmetry through
electroweak sphaleron processes.  The baryon asymmetry is given by
$Y_B \simeq -Y_L/2$, where $Y_L = \kappa \epsilon_1/g^*$, where
$g^* \sim 200$ is the effective number of degrees of freedom in
equilibrium during leptogenesis, and $\kappa$ is the efficiency
factor obtained by solving the Boltzman's equations.  A simple
approximate formula for $\kappa$ is \cite{plu}
\begin{equation}
\kappa \simeq 10^{-2} \left[{0.01 \over \tilde{m}_1~ {\rm
eV}}\right]^{1.1}
\end{equation}
where
\begin{equation}
\tilde{m}_1 = {v_u^2 \over M_1} [\hat{Y}_\nu
\hat{Y}_\nu^\dagger]_{11}
\end{equation}
For $\delta \sim 0.1$ and $M_1 \sim 10^{14}$ GeV, we obtain $Y_B
\sim 7 \times 10^{-11}$, in good agreement with observations.

For the case of inverted mass hierarchy, $N_3$ is lighter than $N_1$, so we focus on $\epsilon_3$.  It
is given by
\begin{equation}
\epsilon_3 \simeq {\mp 3 |Y_\nu^0|^2 \over 4 \pi}\delta^2
\left({m_3 \over m_1}\right) {|x| \sqrt{1-|x|^2/4} \over \sqrt{1+2
|x|^2}}~.
\end{equation}
Again we see that reasonable lepton asymmetry is generated.

In summary, we have presented a class of renormalizable  gauge
models based on the non--Abelian discrete symmetry $A_4$ which
realize the geometric neutrino mixing pattern of Eq.
(\ref{geometric}) naturally. The resulting constraints on the
neutrino masses have been outlined.  We have also highlighted an
intriguing connection between high scale leptogenesis and low
energy neutrino experiments.

\section*{Acknowledgments} The work of KSB is supported in part by
the US Department of Energy grant \#DE-FG02-04ER46140 and
\#DE-FG02-04ER41306.  The work of X-G.H is is supported in part by
a grant from NSC. KSB would like to thank NCTS/TPE at the National
Taiwan University for hospitality where this work was initiated.

\end{document}